\documentclass[a4paper,11pt]{article}
\usepackage{jcappub}

\usepackage{graphicx}
\usepackage{tabularx}
\usepackage[usenames,dvipsnames]{color}
\usepackage{hyperref}
\hypersetup{colorlinks,citecolor= blue,linkcolor= blue, urlcolor=blue}

\usepackage{amsmath}
\usepackage{bbm}
\usepackage{amsfonts}
\usepackage{amssymb}
\usepackage{latexsym}
\usepackage{graphicx}
\usepackage[english]{babel}
\usepackage{multirow}
\usepackage{float}
\usepackage{url}
\usepackage{slashed}
\usepackage{xcolor} 
\usepackage{array}
\usepackage{cases}
\usepackage{paralist}

\newcommand{\be}{\begin{equation}}
\newcommand{\ee}{\end{equation}}
\newcommand{\ba}{\begin{array}}
\newcommand{\ea}{\end{array}}
\newcommand{\bea}{\begin{eqnarray}}
\newcommand{\eea}{\end{eqnarray}}

\usepackage{ulem,fancyvrb}
\usepackage{xcolor}

\title{Distinct photon-ALP propagation modes}

\author[a,b]{Qing-Hong Cao,}
\author[c]{Zuowei Liu,}
\author[a]{Jun-Chen Wang}

\affiliation[a]{School of Physics, Peking University, Beijing 100871, China}
\affiliation[b]{Center for High Energy Physics, Peking University, Beijing 100871, China}
\affiliation[c]{Department of Physics, Nanjing University, Nanjing 210093, China}

\abstract{
Measurement of cosmic photons 
may reveal their propagation 
in the interstellar environment, 
thereby offering a promising way 
to probe axions and 
axion-like particles (ALPs). 
Numerical methods are usually used to compute 
the propagation of the photon-ALP beam due to 
the complexity of both the interstellar magnetic field 
and the evolution equation. 
However, under certain conditions, 
the evolution equation can be greatly simplified  
so that the photon-ALP propagation can  
be analytically solved. 
By using analytic methods, we find 
two distinct photon-ALP propagation modes, 
determined by the relative magnitude of the photon-ALP mixing term 
in comparison to the photon attenuation term.
In one mode, the intensity of photons decreases with the increasing distance; 
in the other mode, it also exhibits oscillatory behavior.
To distinguish the two propagation modes, 
we compute the observable quantities such as  
the photon survival probability 
and the degree of polarization. We also determine through analytic methods 
the conditions under which maximum polarization can be observed and the corresponding upper bound of the survival probability.
}

\emailAdd{qinghongcao@pku.edu.cn}
\emailAdd{zuoweiliu@nju.edu.cn}
\emailAdd{junchenwang@stu.pku.edu.cn}

\begin{document}

\maketitle

\section{Introduction}

The existence of axions and axion-like particles (ALPs) is a topic of great interest in modern particle physics 
\cite{Ringwald:2012hr, Irastorza:2018dyq, Adams:2022pbo}. 
The QCD axion was 
initially proposed as a natural solution to the strong CP problem \cite{Peccei:1977hh,Wilczek:1977pj}. 
Recently, the study of ALPs has gained widespread attention due to the much wider range of mass and coupling parameters than the QCD axion \cite{Jaeckel:2010ni}. ALPs 
arise naturally in a plethora of extensions of the standard model, including supersymmetric models \cite{Chang:1999si,Turok:1995ai} and superstring theories \cite{Svrcek:2006yi,Ringwald:2008cu,Arvanitaki:2009fg}.

The presence of ALPs can significantly alter photon propagation in the universe, leading to distinct photon-ALP oscillation signatures \cite{Maiani:1986md,Raffelt:1987im}.
Because of the complex astrophysical environment through which photon-ALP beams propagate,
most recent studies used numerical simulations to analyze their propagation properties; see e.g., \cite{Li:2020pcn,Horns:2012pp,Gill:2011yp,Bi:2020ths,Galanti:2022yxn,Galanti:2022tow,Xia:2019yud,Liang:2018mqm}. 
Nevertheless, there are also a number of analytical studies on the photon-ALP propagation, 
see e.g., \cite{Raffelt:1987im,Grossman:2002by,Csaki:2003ef,Lai:2006af,DeAngelis:2007dqd,Agarwal:2008ac,Ganguly:2008kh,Mirizzi:2009aj,Wang:2015dil,Kartavtsev:2016doq}. 
Although numerical simulations lead to more accurate results 
than analytical calculations, some 
physical phenomena that can be 
discerned in analytical calculations
are often obscure in numerical simulations.  
In this paper we use analytic methods 
to study the propagation of the photon-ALP beams. 
We find that under certain conditions the evolution equation of the photon-ALP beam 
can be significantly simplified so that analytic methods can be used. 
Our analytic methods reveal two distinct photon propagation modes, 
determined by the relative magnitude of the photon-ALP mixing term 
in comparison to the photon attenuation term.
More precisely, the two propagation modes are 
determined by the sign of the quantity $D$ 
given in Eq.~\eqref{2eq5}:  
in the $D>0$ case,
photons exhibit clear oscillations during the propagation;  
conversely, in the $D<0$ case, the oscillations are absent. 
We investigate the astrophysical conditions that lead to the emergence of these two propagation modes, and further study the implications for relevant physical observables such as the photon survival probability and degree of polarization. 
We study the distinguishing features of the two propagation modes in detail, 
which can be used to discriminate between the two modes, as well as against the standard model background.

The rest of paper is organized as follows. 
In section \ref{PE}, we introduce the ALPs and 
the propagation equation governing the photon-ALP system. 
In section \ref{TWO}, we introduce the two distinct photon propagation modes that are characterized by the sign of a quantity $D$. We compute the density matrix via analytic methods 
and further discuss the effects of ALPs on the density matrix. 
In sections \ref{sec:highE} and \ref{sec:lowE}, 
we discuss the physical observables such as the photon survival probability and the photon degree of polarization 
in the two propagation modes: 
We focus on photons with energies above $100$ GeV 
in section \ref{sec:highE}, 
and photons with energies in the range of $10^{-3}-10^2$ GeV 
in section \ref{sec:lowE}. 
In section \ref{sec:conclusions}, we summarize our findings.
In the appendix, we compare our analytic method in which a uniform 
magnetic field is assumed, 
with numerical calculations in which magnetic field 
models that aim to describe the astrophysical conditions 
are used.

\section{Propagation equation}
\label{PE}

Consider the following effective interaction Lagrangian between the 
axion-like particle (ALP) $a$ and the photon
\begin{equation}
\label{lag}
\mathcal{L}_{\rm int}= 
-\frac{1}{4}g_{a\gamma}F^{\mu\nu}\tilde{F}_{\mu\nu}\, a
=g_{a\gamma}\mathbf{E}\cdot \mathbf{B} \, a,
\end{equation}
where $F^{\mu\nu}$ is the electromagnetic stress tensor and $\tilde{F}_{\mu\nu}$ is the dual, 
$\mathbf{E}$ and $\mathbf{B}$ are the electric and magnetic fields, respectively, 
and $g_{a\gamma}$ is the coupling constant.

The propagation of the photon-ALP beam along 
the $z$-direction with energy $\omega$
can be described by a three-component vector $\psi=(A_x,A_y,a)^T$, 
where $A_{x}$ and $A_{y}$ are the electromagnetic potentials linearly polarized 
along the $x$ and $y$ axes, respectively. 
Without loss of generality, we consider the case where the external magnetic field is along the 
$y$ direction and the equation that governs the propagation of $\psi$ is given by 
\cite{Raffelt:1987im,DeAngelis:2011id,Galanti:2022ijh},  
\begin{equation}
    \label{peq1}
    \left( i\frac{d}{dz} + \omega + \mathcal{M}\right)\psi=0,  
\end{equation}
where $\mathcal{M}$ is given by
\begin{equation}
    \label{peq10}
    \mathcal{M}=\begin{pmatrix}
    \Delta_x & 0 & 0\\
    0 & \Delta_y & \Delta_{a\gamma}\\
    0 & \Delta_{a\gamma} & \Delta_{aa}
    \end{pmatrix}+\frac{1}{2}
    \begin{pmatrix}
    i\Gamma & 0 & 0\\
    0 & i\Gamma & 0\\
    0 & 0 & 0
    \end{pmatrix}.
\end{equation}
Here, $\Delta_x$ and $\Delta_y$ describe the medium effects, 
$\Gamma$ is the absorption rate accounting for the attenuation of photons, 
$\Delta_{a\gamma}=g_{a\gamma}B/2$ is the photon-ALP mixing term 
with $B=|\boldsymbol{B}|$, 
and $\Delta_{aa}=-m_{a}^{2}/(2\omega)$ with $m_{a}$ being the ALP mass.
The absorption rate $\Gamma$ is mainly caused by the reaction $\gamma \gamma \to e^{+}e^{-}$ between the propagating photons and ambient photons, such as cosmic microwave background, extragalactic background light, and so on \cite{Vernetto:2016alq,Lipari:2018gzn}. As shown in the black line of Fig.\ (\ref{fig:delta}), $\Gamma$ becomes significant as energy $\omega \gtrsim 10^5$ GeV.

The medium effects are given by 
\begin{equation}
\Delta_{x,y}= N \Delta_{\rm QED}+\Delta_{\rm pl}+\Delta_{\rm dis},     
\end{equation}
where 
$N=2$ $(7/2)$ for $\Delta_{x}$ ($\Delta_{y}$), 
$\Delta_{\rm QED}$ represents the QED birefringence, 
$\Delta_{\rm pl}$ represents the plasma effect, 
and $\Delta_{\rm dis}$ accounts for dispersion effects from photon-photon scattering on environmental radiation field 
\cite{Bi:2020ths,Galanti:2022yxn}. 
The plasma effect $\Delta_{\rm pl}$ is given by 
\begin{equation}
    \label{peq16}
    \Delta_{\rm pl}=-\frac{\omega_{\rm pl}^2}{2\omega}=-\frac{e^2n_e}{2m_e\omega},
\end{equation}
where $\omega_{\rm pl}$ is the plasma frequency, 
$e$ is the QED coupling, 
and $m_e$ and $n_e$ is the mass and number density of the electron, respectively.  
For the case of $\omega \lesssim 10^5\ \text{GeV}$, 
the QED birefringence and dispersion effects 
can be obtained from the Euler-Heisenberg Lagrangian~\cite{Dittrich:2000zu,Latorre:1994cv,Cougo-Pinto:1999agw,Tarrach:1983cb,Adler:1971wn} 
\begin{align}
    \Delta_{\rm QED}&=\frac{\alpha \omega}{45\pi}\left(\frac{Be}{m_e^2}\right)^2,    \label{peq13}\\
    \Delta_{\rm dis}&=\frac{44\alpha^2\rho_{\rm RF}\omega}{135m_e^4},\label{peq17} 
\end{align}
where $\alpha$ is the fine structure constant, 
and $\rho_{\rm RF}$ is the ambient photon energy density. 
For the high energy cosmic gamma-ray with $\omega \gtrsim \omega_2=10^5$ GeV, 
Euler-Heisenberg approximation breaks down, and one can 
use the a scaling function $g_2(\omega/\omega_2)$ to take into account the gamma-gamma scattering 
due to background photon\ \cite{Dobrynina:2014qba}. 
Thus, at photon energy $\omega \gtrsim \omega_2=10^5$ GeV, the quantities  
$\Delta_{\rm QED}$ and $\Delta_{\rm dis}$ are both modified to 
$\Delta_{\rm QED(dis)}g_2(\omega/\omega_2)$, 
as follows \cite{Dobrynina:2014qba}:
\begin{equation}
\begin{split}
    g_2(u)&=\frac{15}{\pi^4}\int_{0}^{\infty}dx \frac{x^3}{e^x-1}g_1\left(u x \frac{30\zeta(3)}{\pi^4}\right),\\
    g_1(u)&=\frac{3}{8}\int^{+1}_{-1}d\mu (1-\mu)^2g_0\left(u\frac{1-\mu}{2}\right), \\
    g_0(u)&=\frac{45}{44}\int_{1}^{\infty}du'\frac{f(u')}{u'^2-u^2},\\
    f(u)&=\frac{4u(u+1)-2}{u^3}\ln (\sqrt{u}+\sqrt{u-1})-\frac{2(u+1)\sqrt{u-1}}{u^{5/2}}.
\end{split}
\end{equation}

Note that Eq.\ (\ref{peq1}) is similar to the Schr\"odinger equation if the coordinate $z$ is replaced by the time $t$ 
and $(\omega +{\cal M})$ is replaced by $-\mathcal{H}$,
where $\mathcal{H}$ denotes the Hamiltonian. 
Thus, analogous to solving the Schr\"odinger equation, one can also construct the transition matrix $U(z)=e^{i\mathcal{M}z}$ for the photon-ALP propagation.
For a more general case, one can
define the density matrix $\rho \equiv \psi\psi^{\dagger}$, 
which satisfies an equation similar to the Von Neumann equation,
\begin{equation}
    \label{peq3}
    i\frac{d}{dz}\rho=\rho\mathcal{M}^{\dagger}-\mathcal{M}\rho. 
\end{equation}
The solution can then be obtained via 
$\rho(z) = U(z) \rho_0 U(z)^\dagger$, 
where $\rho_0$ is the initial condition. 
In our analysis, we assume an unpolarized photon beam 
such that $\rho_0=\text{diag}(1/2,1/2,0)$.
The upper-left $2\times2$ submatrix of $\rho$, denoted as $\rho_\gamma$, 
can be parameterized as\ \cite{Kosowsky:1998mb}
\begin{equation}
    \label{peq5}
    \rho_{\gamma}=
    \begin{pmatrix}
    \rho_{x} & \rho_{xy}\\
    \rho_{xy}^{*} & \rho_{y}
    \end{pmatrix}
    =\frac{1}{2}
    \begin{pmatrix}
    I+Q & U-iV\\
    U+iV & I-Q
    \end{pmatrix}, 
\end{equation}
where $I$, $Q$, $U$, and $V$ are the Stokes parameters. The photon degree of polarization $\Pi_{L}$ is then given by\ \cite{defin}
\begin{equation}
    \label{peq6}
    \Pi_{L}=\frac{\sqrt{Q^2+U^2}}{I}. 
\end{equation}
The photon survival probability after propagation is given by  \cite{DeAngelis:2011id}
\begin{equation}
    \label{peq8}
    P_{\gamma \to \gamma}=\rho_{x}+\rho_{y}=I.
\end{equation}

\section{Two different propagation modes}
\label{TWO}

Although recent studies focused on numerical methods to 
solve the photon-ALP propagation, due to 
the complex medium effects and  
the magnetic fields in the astrophysical environments, 
there are instances where 
analytic calculations are good approximations to the photon propagation. 
In this section we first identify the conditions in which 
analytic methods to the photon propagation can be used. 
We then discuss two different propagation modes that are 
found in our analysis.

\begin{figure}[t]
  \centering
  \includegraphics[width=0.5\textwidth]{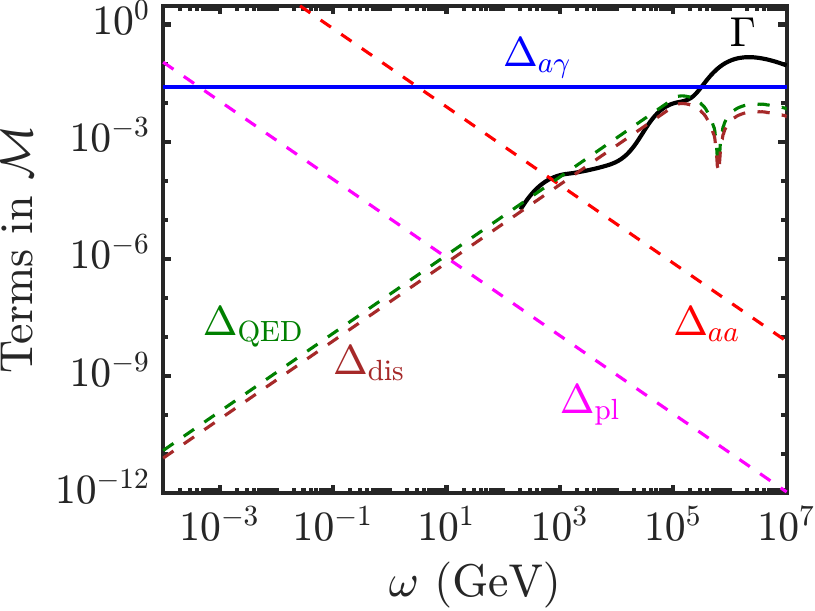}
  \caption{Matrix elements of $\mathcal{M}$ as a function of 
  the photon energy $\omega$. Here we take $g_{a\gamma}=3\times 10^{-12}\ \text{GeV}^{-1}$, $B=5.5\ \mu$G, $m_a=1\times 10^{-18}\ \text{GeV}$, $n_{e}=0.1\ \text{cm}^{-3}$ 
  accounting for the electron number density in the Andromeda galaxy (M31) \cite{mcdaniel2019exploring,Caprioli:2011lfx,Caprioli:2009fv}
  and $\rho_{\rm RF}=2.01\times10^{-51}\ \text{GeV}^{4}$ accounting for the energy density of CMB \cite{Dobrynina:2014qba}. 
We use Eq.\ (\ref{peq13}) and Eq.\ (\ref{peq17}) 
to compute $\Delta_{\rm QED}$ and $\Delta_{\rm dis}$; 
for $\omega>$10$^5$ GeV, 
we further multiply a factor of $g_2(\omega/\omega_2)$ 
where $\omega_2=10^5$ GeV given in Ref.\ \cite{Dobrynina:2014qba}. 
The $\Gamma$ curve is adopted from Ref.\ \cite{Lipari:2018gzn}. 
} 
  \label{fig:delta} 
\end{figure}

We will use the high-energy photons as the 
example in this section, though analytic methods 
can also be used for low-energy photons. 
We first consider the very high energy photons 
with $\omega > 10^5$ GeV.
At such high energy, the matrix $\cal M$ can be greatly 
simplified, because the various $\Delta$ terms can be neglected 
except the new physics term $\Delta_{a\gamma}$ 
and the absorption term $\Gamma$, 
as shown in Fig.\ (\ref{fig:delta}),  
where $g_{a\gamma}=3\times 10^{-12}\ \text{GeV}^{-1}$ and $B=5.5\ \mu$G 
are assumed. 
Thus we have 
\begin{equation}
    \label{2eq1}
    \mathcal{M} \simeq \frac{1}{2}
    \begin{pmatrix}
    i\Gamma & 0 & 0\\
    0 & i\Gamma & g_{a\gamma}B\\
    0 & g_{a\gamma}B & 0
    \end{pmatrix}. 
\end{equation}
We further assume that the variation of external magnetic field is relatively small 
so that it can be approximated by a uniform magnetic field. 
In this case, one can then analytically solve the propagation. 
We note that the analytical solution can facilitate the calculations and 
can reveal some physics pictures of the problem 
that are difficult to be seen in the numerical calculations.

We then find that 
in the analytic solutions for the simplified $\cal M$, 
there exist two different propagation modes of photons,  
characterized 
by the sign of $D$, which is defined as 
\begin{equation}
    \label{2eq5}   
    D \equiv g_{a\gamma}^2B^2 - \frac{\Gamma^2}{4}. 
\end{equation}
We discuss these two propagation modes: 
$D>0$ and $D<0$ below.

\subsection{The $D>0$ propagation mode}

We first discuss the $D>0$ case. 
To solve the propagation analytically, we first compute 
the transition matrix 
$U(z)=e^{i\mathcal{M}z}$, 
where $\mathcal{M}$ is given by Eq.\ \eqref{2eq1}.
Thus, for the $D>0$ case, we have
\begin{equation}
\label{upp}
U_{D>0}(z)=\frac{e^{-\frac{\Gamma}{4}z}}{\cos \varphi}
\begin{pmatrix}
e^{-\frac{\Gamma}{4}z}\cos{\varphi} & 0 & 0\\
0 & \cos \left(\varphi + \frac{\Delta}{2}z \right) & i\sin \frac{\Delta}{2}z \\

0 & i\sin \frac{\Delta}{2}z & \cos \left(\varphi - \frac{\Delta}{2}z \right)
\end{pmatrix},
\end{equation}
where we have defined  
$\Delta=\sqrt{D}$ and $\varphi=\arctan(\Gamma/(2\Delta))$. 
We then compute $\rho(z)$ at distance $z$ from the source via 
$\rho(z) = U(z) \rho_0 U(z)^\dagger$, 
where $\rho_0=\text{diag}(1/2,1/2,0)$ is initial condition  
for an unpolarized photon beam at the source.
The matrix elements of 
$\rho_\gamma(z)$ in this case are given by
\begin{align}
\label{AX1}
\rho_{x}(z) &=\frac{1}{2}e^{-\Gamma z},\\
\label{AY1}
\rho_{y}(z) & =\frac{1}{2} 
\frac{e^{-\frac{\Gamma}{2}z}}{\cos^2 \varphi} \cos^2 
\left(\varphi + \frac{\Delta}{2}z \right),\\
\rho_{xy} (z) &= 0.
\end{align}
Here $\rho_x(z)$ and $\rho_y(z)$ describe the 
intensities of photons polarized along $x$ and $y$ directions, respectively.  
Because the off-diagonal element 
$\rho_{xy} (z)$ vanishes in this case, 
the photon degree of polarization $\Pi_{L}$ 
becomes
\begin{equation}
    \Pi_L = \frac{|\rho_x - \rho_y|}{\rho_x + \rho_y}. 
    \label{eq:PiL}
\end{equation}

We note that the intensities of photons with polarization along the $x$ and $y$ directions exhibit different dependencies on the propagation distance $z$. 
We discuss the distinct behaviors below. 

First, the intensity of photons polarized along the $y$ 
direction depends both on the attenuation term $\Gamma$ 
and on the ALP-interaction term 
$\Delta_{a\gamma}$ (through $\Delta$ and $\varphi$), 
but the intensity along the $x$ 
direction depends on $\Gamma$ only. 
This discrepancy arises from the fact 
that the external magnetic field is taken 
to be along the $y$ direction 
so that photons polarized along 
the $x$ direction are not directly affected by the ALP.

Second, the intensity of photons polarized along the $x$ 
direction decreases exponentially as the distance $z$ increases, 
with a decay length of $L_d=\Gamma^{-1}$, 
as shown in Eq.\ \eqref{AX1}. 
On the other hand, the intensity for photons polarized along the $y$ direction  
appears to exhibit a longer decay length of $L_d=(\Gamma/2)^{-1}$
(twice of that along the $x$ direction), 
based solely on the argument of the exponential function in Eq.\ \eqref{AY1}. 
This change on the decay length is due to the fact that 
photons polarized along the $y$ direction can convert into 
ALPs that do not experience the photon attenuation effects 
$\Gamma$. 
We emphasize that the actual decay length 
for photons polarized along the $y$ direction will deviate 
somewhat from the value of $L_d=(\Gamma/2)^{-1}$ due to the 
additional $z$-dependence in Eq.\ \eqref{AY1}. 
Specifically, taking the limit $g_{a \gamma} \to 0$ in Eq.\ \eqref{AY1} 
introduces an additional exponential factor $e^{-(\Gamma/2) z}$, which must be taken into account. 
\footnote{Note that the $g_{a \gamma} \to 0$ limit is not allowed in the $D>0$ case. 
See more discussions in the $D<0$ case.}

Third, 
the intensity of photons polarized along the $y$ direction, 
while propagating, exhibits an oscillatory behavior. 
We define the oscillation length, denoted by $L_o$, as 
the period of the absolute value of the cosine function 
in Eq.\ \eqref{AY1}. 
The oscillation length is given by
\begin{equation}
    L_o = \frac{2 \pi}{\Delta} = \frac{2 \pi}{\sqrt{g_{a \gamma}^2B^2-\left(\frac{\Gamma}{2}\right)^2}}.  
\end{equation}
Thus, the increase in the difference between $g_{a \gamma} B$ and $\Gamma$ leads to an increase in the oscillation length. We note that the oscillation is attenuated by the exponential factor $e^{-(\Gamma/2)z}$.

\subsection{The $D<0$ propagation mode}

We next discuss the $D<0$ case. 
For the case of $D<0$, the transition matrix $U(z)$ can be written as
\begin{equation}
\label{ump1}
U_{D<0}(z)=
\frac{e^{-\frac{\Gamma}{4}z}}{\cos 2\alpha}
\begin{pmatrix}
e^{-\frac{\Gamma}{4}z}\cos 2\alpha & 0 & 0\\
0 & e^{-\frac{\Delta}{2}z}\cos ^2 \alpha - e^{\frac{\Delta}{2}z}\sin ^2 \alpha & i \left(e^{\frac{\Delta}{2}z}-e^{-\frac{\Delta}{2}z}\right)\sin \alpha \cos \alpha \\
0 & i \left(e^{\frac{\Delta}{2}z}-e^{-\frac{\Delta}{2}z}\right)\sin \alpha \cos \alpha & e^{\frac{\Delta}{2}z}\cos ^2 \alpha - e^{-\frac{\Delta}{2}z}\sin ^2 \alpha
\end{pmatrix},
\end{equation}
where $\Delta=\sqrt{-D}$
and $\alpha=\arcsin {\sqrt{\Gamma/2-\Delta}}/{\sqrt{\Gamma}}$. 
\footnote{The definition of $\Delta$ in the $D<0$ case is different from the $D>0$ case.}

For an unpolarized photon beam at the source, 
the diagonal matrix elements of 
$\rho_\gamma(z)$ are given by
\begin{align}
\label{AX2}
\rho_{x}(z) &=\frac{1}{2}e^{-\Gamma z},\\
\label{AY2}
\rho_{y}(z) &=\frac{1}{2}\frac{e^{-\frac{\Gamma}{2}z}}{\cos^2 (2\alpha)}
e^{-{\Delta}z}
\left|\cos ^2 \alpha - e^{\Delta z}\sin ^2 \alpha \right|^2, \\
\rho_{xy} (z) &= 0. 
\end{align}

We note that both $\rho_{x}(z)$ and $\rho_{xy}(z)$ 
are the same as in the $D>0$ case. 
Similarly to the $D>0$ case, the vanishing off-diagonal element $\rho_{xy}(z)$ leads to a simplified expression of the photon 
as given in Eq.~\eqref{eq:PiL}. 
We next discuss the distinct dependencies on the 
propagation distance $z$ of the intensities of photons 
with polarization along the $x$ and $y$ directions, 
and also make comparison to the $D>0$ case.

First, photons polarized along the $x$ direction only 
undergo attenuation characterized 
by the photon attenuation coefficient $\Gamma$, 
whereas photons polarized along the $y$ direction  
are influenced both by $\Gamma$ 
and by the mixing term with the ALP. 
This is similar to the $D>0$ case.

Second, the ALP-photon mixing term weakens the photon attenuation. 
This can be seen from the argument of the exponential function 
in Eq.~\eqref{AY2}, 
which hints a decay length of $L_d = (\Gamma/2+\Delta)^{-1}$ 
for $y$-polarized photons. 
The decay length of $x$-polarized photons is $L_d = \Gamma^{-1}$. 
Since $\Delta < \Gamma/2$, 
photons polarized along the $y$-direction has a larger 
decay length than photons polarized along the $x$-direction. 
The reason behind this is the same as the $D>0$ case: 
when propagating, $y$-polarized photons can convert into ALPs 
which are unaffected by photons attenuation effects.

Third, in contrast to the $D>0$ scenario, where photons polarized in the $y$ direction exhibit oscillations, photons in the $D<0$ scenario do not exhibit any oscillatory behavior.

\section{Propagation modes for high-energy photons with $\omega \geq 100$ GeV}
\label{sec:highE}

In this section 
we discuss the propagation modes 
for photons with energy $\omega \geq 100$ GeV. 
\footnote{Note that in the energy range of $10^{3}$ GeV $\lesssim \omega \lesssim 10^5$ GeV, 
$\Delta_{\rm QED(dis)} > \Gamma$ so that one can no longer keep $\Gamma$ 
while neglecting $\Delta_{\rm QED(dis)}$. 
However, for sufficiently large $g_{a\gamma}B$ values, both 
$\Gamma$ and $\Delta_{\rm dis}$ can be neglected, leading to 
a simpler expression than Eq.~(\ref{2eq1}); 
see more discussions on this case in section \ref{sec:lowE}.}
In the Milky Way (MW) and many other spiral galaxies including 
the Andromeda galaxy (M31), 
the strength of the magnetic field is of 
$O(1)\ \mu$G \cite{Jansson:2012pc,Jansson:2012rt,Fletcher:2003ec}. 
Taking $B=1\ \mu$G and  
$g_{a\gamma}=3\times 10^{-12}$ GeV$^{-1}$, 
we find that 
$B g_{a\gamma} \simeq 0.01$ kpc$^{-1}$; 
equating $B g_{a\gamma}$ with $\Gamma/2$ leads to 
the photon energy at $\sim 250$ TeV. 
Thus, in this case,
the photon propagation mode is the $D<0$ mode 
for $\omega\gtrsim 250$ TeV, 
and the $D>0$ mode for $\omega\lesssim 250$ TeV. 
For active galactic nuclei, galaxy clusters, and intergalactic space, the typical magnetic field strengths are $O(1-10^{5})\ \mu$G \cite{Pudritz:2012xj,Tavecchio:2009zb}, $O(1-10)\ \mu$G \cite{Govoni:2004as}, and $O(10^{-7}-1)$ nG \cite{Neronov:2010gir,Pshirkov:2015tua}, respectively. 
Because the magnetic field in active galactic nuclei 
and galaxy clusters can be much larger than that in spiral galaxies, one can have the $D>0$ mode 
for photons with energy $\omega \geq 100$ GeV. \footnote{However, if the variation of external magnetic field is large, the mean value of $B$ may be very small and the propagation mode would be more similar to the $D<0$ mode. See Fig.\ (\ref{GC}) for details.}
On the other hand, the magnetic field in 
intergalactic space is much smaller than 
that in the MW galaxy, 
leading to the $D<0$ mode for photons with energy $\omega \geq 100$ GeV. 
We note that our general discussions here are not applicable in extreme environments where the magnitude of the magnetic field and/or electron number density is significantly large such that the terms $\Delta_{\rm QED}$ and/or $\Delta_{\rm pl}$ in the $\mathcal{M}$ matrix can become substantial.

\subsection{Photons from M31}
\label{sec:M31 description}

To be concrete, in this section we focus on 
photons originating from the M31 galaxy and propagating to Earth. 
\footnote{In Appendix \ref{app:Appendix}, 
we also carry out analysis to compare analytic calculations 
with the numerical calculations both in the MW galaxy 
and in galaxy clusters.}
In this case, the photon-ALP propagation consists of three components: 
propagation in the M31 galaxy, 
propagation in the Milky Way (MW) galaxy, and 
propagation through the intergalactic space between the two galaxies.
Note that since the inclination of the M31 galaxy 
is $77.5^{\circ}$, 
\footnote{Note that a galaxy with 
an inclination of $90^{\circ}$ is an edge-on galaxy.}
and the M31 disk is about 1 kpc thick 
\cite{ma1997thickness,hu2013disk}, 
photons propagating in the M31 disk with distance of $\sim 5$ kpc can point to Earth.
Also note that because the M31 galaxy is located at 
RA 00h 42m 44.3s, Dec $41^{\circ}$ 16' 9'', 
photons originating from it have a rather small propagation distance in the MW disk. 
Moreover, the out-of-disk component of the magnetic field in the MW galaxy 
is very small \cite{Jansson:2012pc}. 
Thus, in our analysis we only consider photon-ALP propagation in the M31 galaxy 
and in the intergalactic space, but neglect the propagation in the MW galaxy.

Although the M31 galaxy is the nearest major galaxy to the MW galaxy, 
its high inclination angle presents some challenges for observing its structure 
\cite{arp1964spiral,ferguson2002evidence,Beck:2013bxa}. 
The M31 galaxy comprises several major components, including a disk and a bulge. 
For the bulge radius, various measurements and analyses have produced a wide range of values, 
ranging from 0.1 kpc to 10 kpc \cite{leahy2023complex}. 
Accounting for the uncertainties, we model the M31 galaxy 
with a spherical bulge with radius $r<$ 6 kpc and 
a disk with radius $6 < r < 20$ kpc \cite{hammer20182}. 
We further note that the current understanding of the magnetic field in the M31 galaxy remains limited 
\cite{mcdaniel2019exploring,giessubel2014magnetic}. 
Thus, for the bulge, we adopt a magnetic field of 6 $\mu$G \cite{beck2011cosmic,Beck:2013bxa}, 
and assume that the direction of the magnetic field is azimuthal, analogous to the MW bulge \cite{Jansson:2012pc}. 
For the M31 disk, we only consider the regular component of the magnetic field, 
and adopt the model in Ref.~\cite{Fletcher:2003ec}, 
which is given in Table \ref{M31}. 
We neglect the turbulent field of the M31 galaxy, since its coherence length 
is usually much smaller than the 
$\gamma \leftrightarrow a$ oscillation length \cite{Galanti:2022yxn}.

\begin{table}[htbp]
\centering
\begin{tabular}{|c|c|c|c|c|}
\hline
   $r$ (kpc) & 6-8  &  8-10 & 10-12 & 12-14 \\
\hline
   $B$ ($\mu$G) & 4.9 &   5.2 & 4.9 & 4.6  \\
\hline
\end{tabular}
\caption{The regular component of the magnetic field in the M31 disk, 
adopted from Ref.~\cite{Fletcher:2003ec}.
Here $r$ is the radial coordinate and $B$ is the magnetic field.}
\label{M31}. 
\end{table}

The magnitude of the intergalactic magnetic field 
has been found  to be in the  range  of 
$3 \times 10^{-7}$ nG $\lesssim$ B $\lesssim 1.7$ nG \cite{Neronov:2010gir,Pshirkov:2015tua}. 
Note that the intergalactic magnetic field is believed to be domain-like 
with a domain length of $\mathcal{O}(1)$ Mpc: 
within one domain, the magnetic field is constant \cite{Kronberg:1993vk,Grasso:2000wj}. 
Since the M31 galaxy is $\sim 765$ kpc away from us, 
in our analysis we assume the intergalactic space between the M31 and MW galaxies 
to be within one domain of the intergalactic magnetic field. 
We thus take a constant magnetic field of $B=1$ nG 
for the intergalactic magnetic field between the M31 and MW galaxies.

To illustrate the two different propagation modes 
($D>0$ and $D<0$), we consider the ALP model where 
$m_a=10^{-18}$ GeV 
\footnote{Note that the analysis is insensitive to the precise value of 
the ultralight ALP mass 
as long as the $\Delta_{aa}=-m_{a}^{2}/(2\omega)$ term is small 
compared to other terms in the $\mathcal{M}$ matrix in Eq.~(\ref{peq10}).} 
and $g_{a\gamma} = 3 \times 10^{-12}$ GeV$^{-1}$, and 
monochromatic photons originating from the M31 galaxy 
with energy of 
$5 \times 10^2$ GeV and $5 \times 10^5$ GeV:
\begin{itemize}

\item For the $5 \times 10^2$ GeV case, we consider
photons originating from the center of the M31 galaxy 
and propagating to Earth such that 
the propagation distances in the M31 galaxy are 
6 kpc in the bulge and 5 kpc in the disk. 
Note that for the $\omega = 5 \times 10^2$ GeV case, 
the attenuation term is $\Gamma \sim 10^{-4}$ kpc$^{-1}$, and  
the photon-ALP mixing term in the intergalactic space 
is $g_{a\gamma}B \sim 10^{-4}$ kpc$^{-1}$,  
where we have used $B = 1$ nG. 
In the contrast, the photon-ALP mixing term in the M31 galaxy is $g_{a\gamma}B \sim  10^{-1}$ kpc$^{-1}$, 
where we have used $B = 6 \, \mu$G. 
Thus, in this case we neglect the propagation in the intergalactic space 
and only consider the propagation in the M31 galaxy.  
Note that the propagation in the M31 galaxy is the $D>0$ mode, 
since $g_{a\gamma}B>\Gamma/2$ in this case.

\item For the $5 \times 10^5$ GeV case, we consider photons originating from the edge of the M31 galaxy so that the propagation within the 
M31 galaxy can be neglected. We thus only need to consider the propagation in the 
intergalactic space.
Note that the attenuation effect for the $\omega = 5\times 10^5$ GeV case 
in the intergalactic space is significant: $\Gamma \sim 10^{-2}$ kpc$^{-1}$, 
which exceeds the photon-ALP term, $g_{a\gamma}B \sim 10^{-4}$ kpc$^{-1}$, in the intergalactic space.
Therefore, the photon propagation in this case 
corresponds to the $D<0$ mode. 

\end{itemize}

\subsection{The $D>0$ case}
\label{sec:Dpositive}

We first discuss the physics in the $D>0$ case, 
including  
the photon degree of polarization 
and the photon survival probability.

\subsubsection{Photon degree of polarization}

For the $D>0$ mode, we first compute 
the photon degree of polarization $\Pi_{L}$, 
which is given by Eq.\ \eqref{eq:PiL}, 
in the absence of the off-diagonal elements of the 
$\rho_\gamma$ matrix. 
Due to the interaction with the ALP, 
photons observed at Earth can be fully polarized, i.e., $\Pi_L=1$, 
in spite of the unpolarized initial condition at the source 
position. 
This is a remarkable signature for ALP detection. 
Thus, it is of great interest to determine under what conditions 
photons observed can be fully polarized. 
The nearly fully-polarized $\Pi_L \sim 1$ can be achieved 
in two cases: 
(1) $\rho_x \gg \rho_y$, and  
(2) $\rho_x \ll \rho_y$.

\begin{itemize}

\item The $\rho_x \gg \rho_y$ case:
Since $\rho_x(z) > 0$, the full-polarization case $\Pi_L=1$ 
can be achieved if $\rho_y(z) = 0$; 
by using the analytic expression of $\rho_y(z)$ 
given in Eq.\ \eqref{AY1}, we obtain the $z$ values 
for $\Pi_L=1$:
\begin{equation}
    \label{z0L}
    z_n=\frac{2}{\Delta}\left[\frac{n-1}{2}\pi+\arctan\left( \frac{2\Delta}{\Gamma}\right)\right],
\end{equation}
where $n$ is a positive odd integer. 
Thus, the full-polarization $\Pi_L=1$ phenomena are  
evenly distributed in space, 
and the distance between adjacent $\Pi_L$ points  
is $2\pi/\Delta$. 
Because $\rho_x(z) > 0$ decreases with $z$, 
the detection should be performed with the 
smallest distance $z_1$ in Eq.\ \eqref{z0L}, 
if possible. 
Note that $z_1=2\arctan(2\Delta/\Gamma)/\Delta$
can be significantly small if 
$g_{a\gamma}B$ is sufficiently large. 

\item The $\rho_x \ll \rho_y$ case: 
Because the $y$-polarized photons have a longer 
decay length than the $x$-polarized photons, 
eventually $\rho_x$ can become much smaller 
than $\rho_y$, 
leading to a relatively large polarization, 
$\Pi_L \simeq 1$. 
By using the 
analytic expressions given in 
Eq.\ \eqref{AX1}
and 
Eq.\ \eqref{AY1}, 
we obtain the maximum values of $\Pi_L$ 
in the $\rho_x \ll \rho_y$ case
\begin{equation}
\label{eq:piLmax2}
     \Pi_L^n = \tanh \left(\frac{\Gamma}{4}z_n\right), 
     \,\, \text{with} \,\,
     z_n = \frac{\pi n}{\Delta}, 
\end{equation}
where $n$ is a positive even integer.
We further extrapolate 
the maximum value of $\Pi_L$ on $z_n$, 
as given in Eq.\ \eqref{eq:piLmax2}, 
to all $z$ values,
\begin{equation}
   \label{AppPi}
    \Pi_L = \tanh \left(\frac{\Gamma}{4}z\right),
\end{equation}
which can be interpreted as the theoretical upper 
bound of $\Pi_L$\textcolor{blue}. 
It is interesting to note that Eq.\ \eqref{AppPi} 
is independent of the external magnetic field. 

\end{itemize}

\begin{figure}[t]
  \centering
  \includegraphics[width=0.45\textwidth]{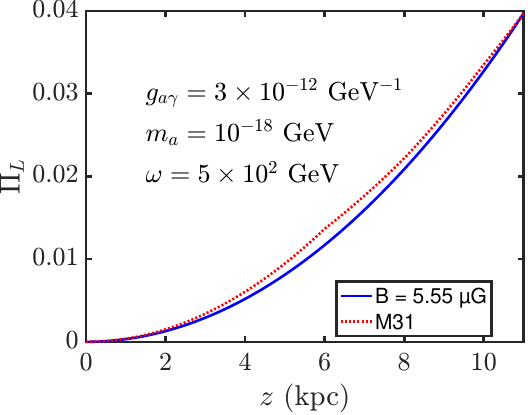}
  \includegraphics[width=0.45\textwidth]{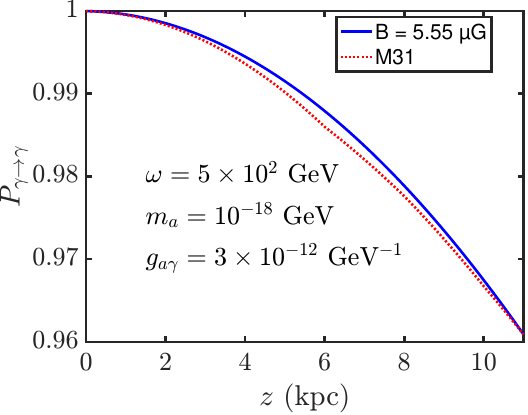}
\caption{The photon polarization degree $\Pi_L$ (left) 
  and the photon survival probability $P_{\gamma\to\gamma}$ (right) 
  as a function of the propagation distance $z$ in the M31 galaxy 
  (red-dotted), in the $D>0$ case. 
  Here we consider photons originating from the 
  central region of the M31 galaxy such that 
  the photon propagation distances are
6 kpc in the bulge and 5 kpc in the disk of the M31 galaxy, respectively.
  Here, we use $g_{a\gamma}=3\times10^{-12}\ \text{GeV}^{-1}$ 
  and $\omega = 500$ GeV. 
  Also shown are the calculations of $\Pi_L$ and $P_{\gamma \to \gamma}$ 
  using a uniform magnetic field $B=5.5\ \mu$G (blue-solid).} 
  \label{fig:M31Dp0} 
\end{figure}

\begin{figure}[htbp]
  \centering
  \includegraphics[width=0.45\textwidth]{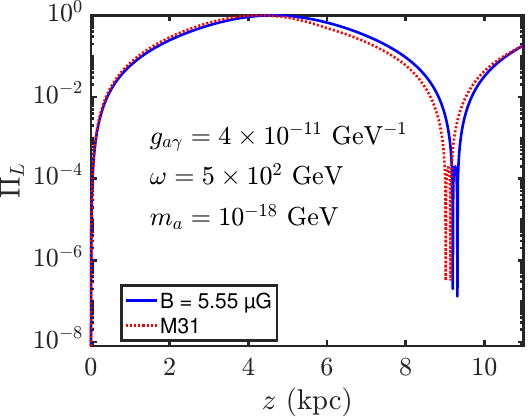}
  \includegraphics[width=0.45\textwidth]{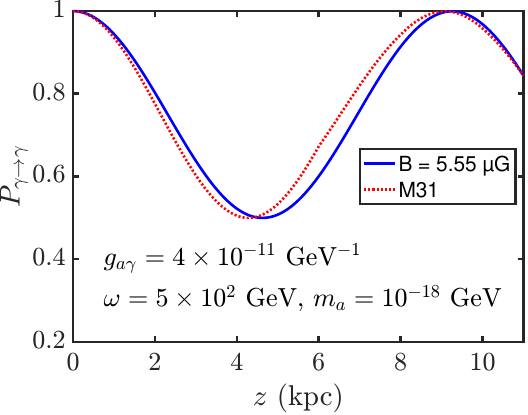}
  \caption{Same as Fig.~\ref{fig:M31Dp0} except  
  $g_{a\gamma}=4\times10^{-11}\ \text{GeV}^{-1}$.}
  \label{fig:M31DpLV} 
\end{figure}

The left-panel figure of Fig.~(\ref{fig:M31Dp0}) 
shows the photon polarization degree $\Pi_L$ 
as a function of the propagation distance $z$, 
for the case where $m_a=10^{-18}$ GeV,  
$g_{a\gamma} = 3 \times 10^{-12}$ GeV$^{-1}$ and $\omega = 500$ GeV.
Note that $\Pi_L$ at different $z$ values can be 
interpreted as the observed polarization for photons with shorter propagation distances in the M31 galaxy. In Fig.~(\ref{fig:M31Dp0}), we analyze photon propagation using two different treatments on the magnetic fields: 
(1) the magnetic model of the M31 galaxy as 
described in section \ref{sec:M31 description}, 
and (2) a constant magnetic field of $B=5.5\ \mu$G, 
which is the average M31 magnetic field.   
We find that the approximation of using  
a constant magnetic field of $B=5.5\ \mu$G 
closely matches the actual magnetic model of the M31 galaxy. Thus, computing photon propagation using the 
analytical formulas in section \ref{TWO} with 
a constant magnetic field is a valid approximation 
for the analysis.

In Fig.~(\ref{fig:M31Dp0}), 
the maximal photon degree of polarization should first occur 
in the $\rho_x \gg \rho_y$ case.
According to Eq.\ (\ref{z0L}), $z_1 \simeq 61.7$ kpc for $n=1$,  
and the distance between adjacent peaks 
in the $\rho_x \gg \rho_y$ case is $2\pi/\Delta \simeq 123.4$ kpc.
We observe that $z_1$ and $2\pi/\Delta$ are not visible in the left-panel figure of Fig.~(\ref{fig:M31Dp0}), 
because both the propagation distance $z \sim 10$ kpc 
and the ALP coupling term  $g_{a\gamma}B \sim 10^{-1}$ kpc$^{-1}$ are relatively small. 
To observe polarization peaks, the condition $g_{a\gamma}B z \gtrsim \mathcal{O}(10)$ must be met. 
Therefore, with ALP parameters $g_{a\gamma}=3\times 10^{-12}$ GeV$^{-1}$ and $m_a = 10^{-18}$ GeV, 
the polarization peaks are unobservable for photons originating from M31. 

To illustrate the underlying physics, we increase the ALP-photon coupling 
to $g_{a\gamma} = 4 \times 10^{-11}$ GeV$^{-1}$ in the left-panel figure of Fig.~(\ref{fig:M31DpLV}). 
The first polarization peak appears at $z=4.63$ kpc, 
corresponding to $z_1$ in Eq.\ \eqref{z0L}. 
Additionally, there also exists a small polarization peak at $z=9.26$ kpc, 
corresponding to $z_2$ in Eq.\ \eqref{eq:piLmax2}.
While the coupling constant $g_{a\gamma} = 4 \times 10^{-11}$ GeV$^{-1}$ 
exceeds current experimental limits, 
it allows the polarization peaks to become clearly visible. 
Alternatively, the desired polarization peaks could also become observable with smaller, 
experimentally allowed ALP couplings 
if the magnetic field is significantly large with a substantial coherence length. For example, in environments such as galaxy clusters where $g_{a\gamma} B z \sim \mathcal{O}(100)$, the required conditions may arise. If such systems are identified in future observations, the same physics demonstrated in Fig.~(\ref{fig:M31DpLV}) could be realized using parameters consistent with current experimental constraints.

\subsubsection{Photon survival probability}

The right-panel figure of Fig.\ (\ref{fig:M31Dp0}) 
shows the photon survival probability 
$P_{\gamma\to\gamma}$ as a function of 
the propagation distance $z$, for the same 
parameters as the left-panel figure of Fig.\ (\ref{fig:M31Dp0}). 
Again, the oscillatory behavior in the photon survival probability due to 
the presence of the ALP is difficult to detect in Fig.\ (\ref{fig:M31Dp0}). 
To illustrate the oscillatory behavior, 
we increase the ALP-photon coupling 
to $g_{a\gamma} = 4 \times 10^{-11}$ GeV$^{-1}$ in the right-panel figure of Fig.~(\ref{fig:M31DpLV}), 
where the photon survival probability reaches its minimum value at $z\simeq 4.63$ kpc. 
Note that the photon survival probability $P_{\gamma\to\gamma}$ 
is bounded from above by 
$P_{\gamma\to\gamma} < (e^{-\Gamma z}+e^{-\Gamma z/2}/\cos \phi)/2$, 
which is larger than the case without ALP, 
$P_{\gamma\to\gamma}=e^{-\Gamma z}$. 
Thus, the presence of ALP makes distant 
galaxies brighter, resulting in a better detection probability.

\subsection{The $D<0$ case}
\label{sec:Dnegative}

\begin{figure}[t]
\centering
\includegraphics[width=0.45\textwidth]{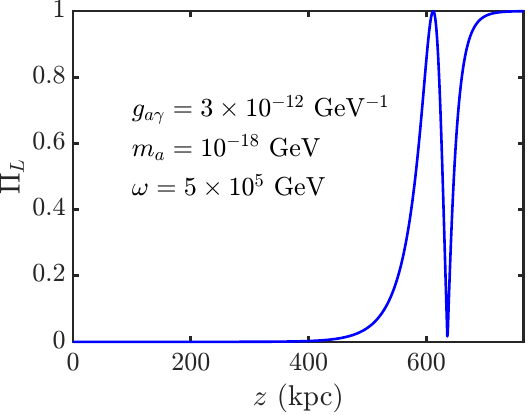}
\includegraphics[width=0.45\textwidth]{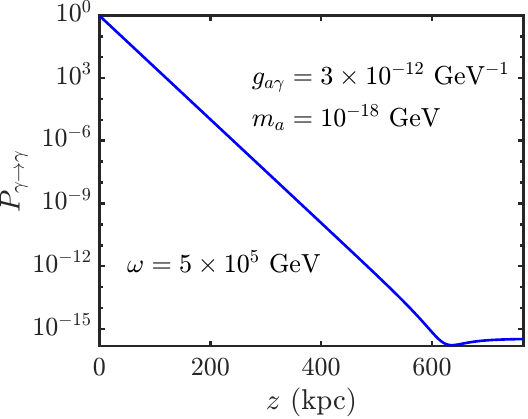}
\caption{The photon polarization degree $\Pi_L$ (left) 
and the photon survival probability $P_{\gamma\to\gamma}$ (right) 
as a function of the photon propagating distance $z$, 
in the $D<0$ case. 
A constant external magnetic $B=1$ nG, 
the coupling constant $g_{a\gamma}=3\times 10^{-12}\ \text{GeV}^{-1}$, 
and the energy $\omega=5\times 10^5$ GeV are used.}
\label{fig:M31Dm0} 
\end{figure}

We next discuss the physics in the $D<0$ case. 
The left panel figure of 
Fig.\ (\ref{fig:M31Dm0}) shows 
the degree of polarization $\Pi_{L}$, for the $\omega = 5\times 10^5$ GeV case, 
as a function of the propagation distance $z$, 
where 
a constant intergalactic magnetic field $B=1\ \text{nG}$, $g_{a\gamma}=3\times 10^{-12}\ \text{GeV}^{-1}$, 
and $\Gamma=0.057\ \text{kpc}^{-1}$ are used.
Similarly to the $D>0$ propagation mode, 
the nearly fully-polarized $\Pi_L \sim 1$ 
in the $D<0$ propagation mode can be also achieved 
in two cases: 
(1) $\rho_x \gg \rho_y$, 
(2) $\rho_x \ll \rho_y$.

We first discuss the $\rho_x \gg \rho_y$ case. 
In this case, 
the full-polarization cases $\Pi_L=1$ 
occurs when $\rho_y(z) = 0$, since  $\rho_x(z) > 0$. 
Unlike the infinite $z$ values for $\rho_y(z) = 0$ in the 
$D>0$ case, there is only a single point 
for $\rho_y(z) = 0$ in the $D<0$ case. 
By using the analytic expression of $\rho_y(z)$ 
in Eq.\ \eqref{AY2}, we obtain the single $z$ value 
for $\rho_y(z) = 0$ as
\begin{equation}
\label{z0s}
z_0=\frac{1}{\Delta}\ln{ \frac{\Gamma+2\Delta}{\Gamma-2\Delta}=\frac{2}{\sqrt{\left(\frac{\Gamma}{2}\right)^2-g_{a\gamma}^2B^2}}\ln{\frac{\frac{\Gamma}{2}+\sqrt{\left(\frac{\Gamma}{2}\right)^2-g_{a\gamma}^2B^2}}{g_{a\gamma}B}}}. 
\end{equation}
In Fig.\ (\ref{fig:M31Dm0}), one has that $z_0 \simeq 612$ kpc, 
which is the first point where the full-polarization cases $\Pi_L$ reaches one.

We next discuss the $\rho_x \ll \rho_y$ case. 
Because the $y$-polarized photons have a longer 
decay length than the $x$-polarized photons, 
eventually $\rho_x$ can become much smaller 
than $\rho_y$, 
leading to a nearly full-polarization, 
$\Pi_L \simeq 1$. 
This occurs 
at a relatively large $z$ value. 
Note that $\rho_y$ has a local maximal point 
at $z = 2 z_0$. 
Thus, we take the $z \gtrsim 2 z_0$ as 
the region for the nearly full-polarization, 
$\Pi_L \simeq 1$.

The right panel figure of Fig.\ (\ref{fig:M31Dm0}) shows 
the photon survival probability $P_{\gamma\to\gamma}$, 
in the $D<0$ case. 
Unlike the $D>0$ case, here $P_{\gamma\to\gamma}$ does 
not oscillate with the propagation distance $z$. 
Nevertheless, the presence of ALPs can still facilitate 
the detection of remote photons more effectively than 
in the absence of ALPs.

We next investigate the optimal conditions 
for observing polarization signals 
in the $D<0$ propagation mode. 
In order to observe a distinct polarization signature with a 
strong signal strength, two conditions must be met. 
Firstly, the degree of polarization should be close to one, 
namely $\Pi_{L}\approx 1$. 
This condition can be achieved either at $z\simeq z_0$
or at $z \geq 2 z_0$. 
Secondly, the total intensity of photons should be significant, 
which implies a significant photon survival probability 
$P_{\gamma\to\gamma}=\rho_x+\rho_y$.  
Since $\rho_x(z)$ decreases with increasing $z$, 
observation at $z=z_0$ should be perused as the first option.

We next study the $z \geq z_0$ region. 
In this region, 
while the intensity $\rho_x(z)$ continues to decrease with increasing $z$, 
the intensity $\rho_y(z)$ starts to grow 
from zero at $z=z_0$ to its local maximum point at $z=2z_0$, 
and then decreases with increasing $z$. 
Therefore, the maximum value of the photon survival probability 
in the $z \geq z_0$ region 
should occur in the interval of $z_0 \leq z \leq 2z_0$; 
within this interval, 
the maximum value of $\rho_x$ is $e^{-\Gamma z_0}/2$ 
at $z=z_0$, which is the same as the maximum value of $\rho_y$ 
at $z= 2 z_0$. 
The theoretical upper bound on the photon survival probability 
can be obtained by simply summing these two maximum values, 
leading to $P_{\gamma \to \gamma} < e^{-\Gamma z_0}$. 
\footnote{The upper bound on the photon survival probability 
obtained in this way is larger than or equal to its maximum value.} 
We next study the possible maximum value of the 
above upper bound. 
By using Eq.\ (\ref{z0s}), we have
$
 \Gamma z_0=2x\ln[(x+1)/(x-1)],
$
where $x \equiv {\Gamma}/(2\Delta)\geq1$. 
When $x\to\infty$, the quantity $\Gamma z_0$ 
reaches its minimum value, which is 4. 
This can be achieved by 
setting $\Delta = 0$, which arises if $\Gamma/2=g_{a\gamma}B$. 
This then leads to a maximum value   
for the upper bound on the photon survival probability 
$P_{\gamma\to\gamma} < e^{-4} \simeq 1.8\%$.

Therefore, for $D<0$ propagation mode, 
the observations on the 
degree of polarization $\Pi_L$ should be first conducted 
at the distance of $z \simeq z_0$ and $z \simeq 2 z_0$. 
The theoretical upper bound on the photon 
survival probability in the $z \geq z_0$ region is 
$\simeq 1.8\%$. However, we note that in Fig.~(\ref{fig:M31Dm0}), the photon survival probability
$P_{\gamma \to \gamma} \simeq 10^{-15}$ 
for $z \geq z_0$;  
thus, it is difficult to observe 
a significant polarization signal in this case.

\begin{figure}[t]
  \centering
  \includegraphics[width=0.45\textwidth]{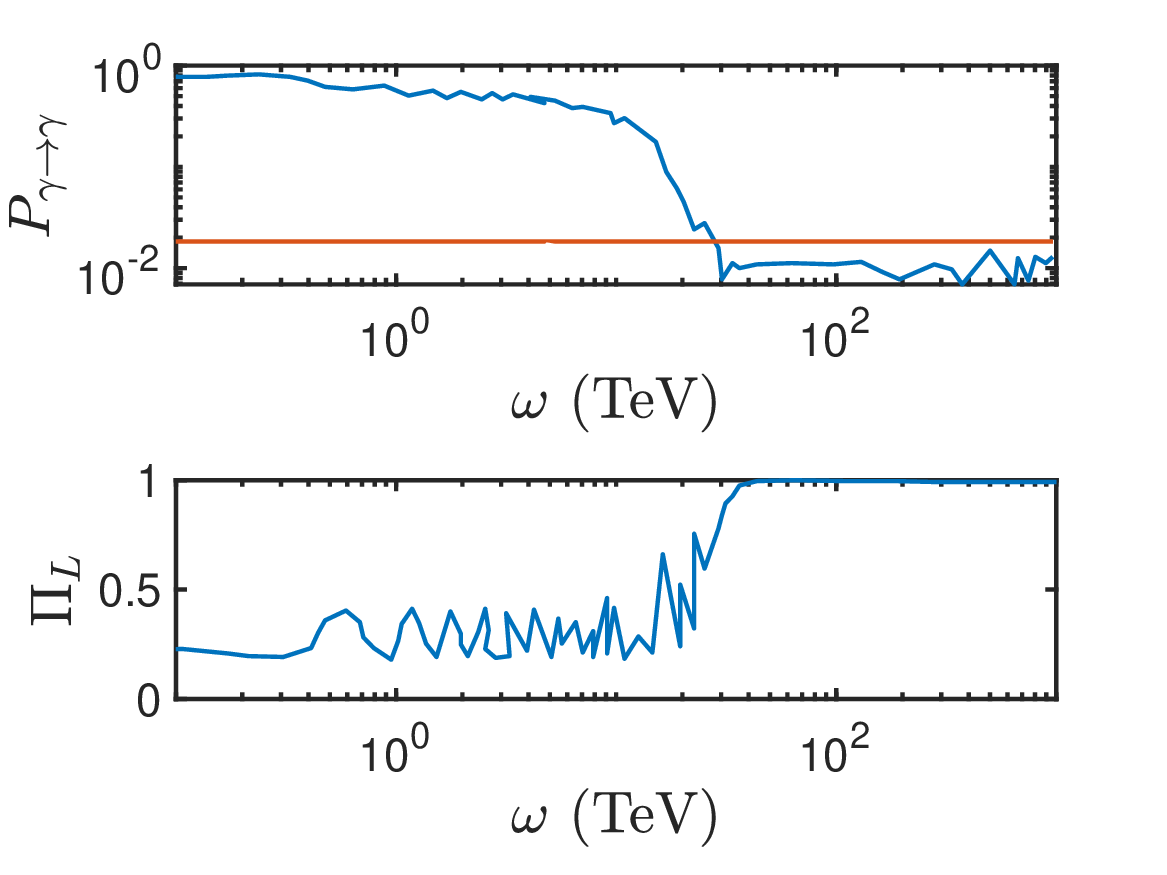}
  \includegraphics[width=0.45\textwidth]{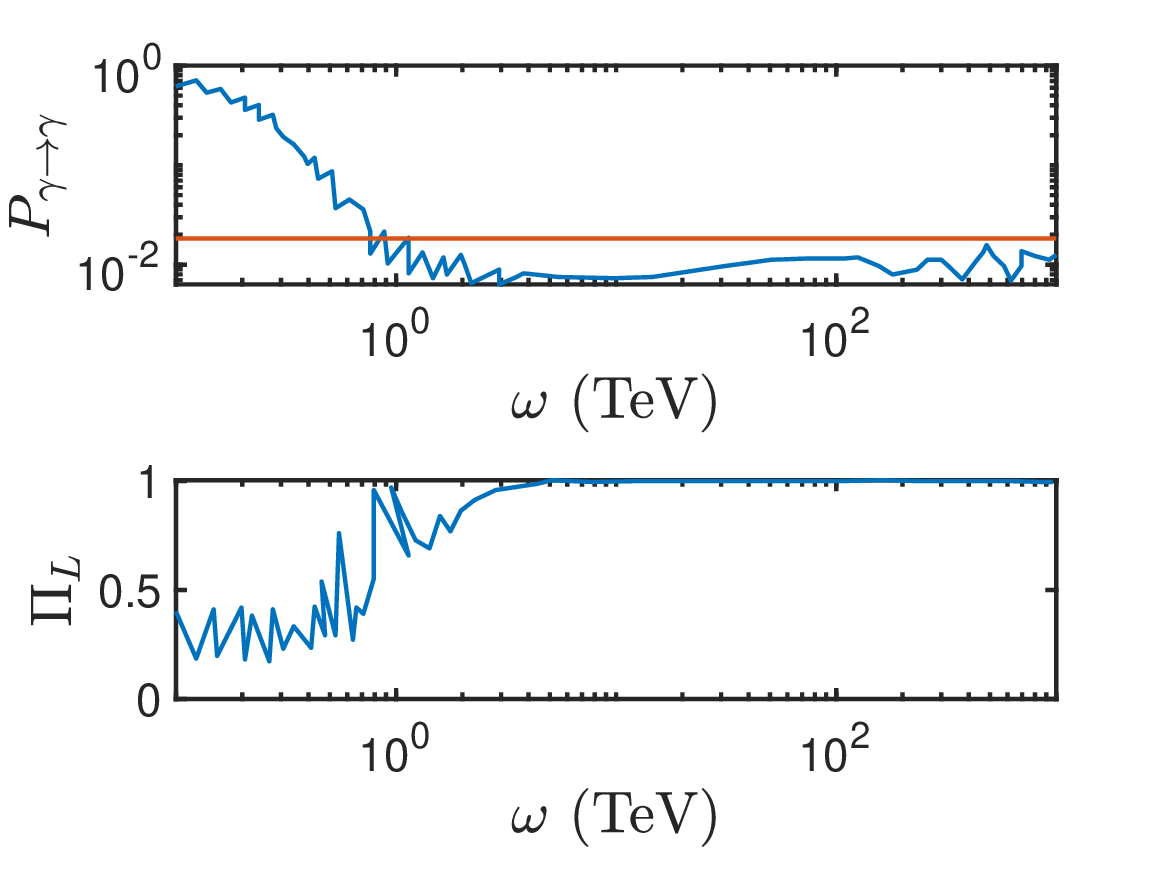}
  \caption{The photon survival probability $P_{\gamma \to\gamma}$ (upper) and the photon degree of polarization $\Pi_{L}$ (lower) as a function 
  of the photon energy. 
  Blue lines are results adopted from 
  Figure 11 of Ref.\ \cite{Galanti:2022yxn}, 
  where photons originate from clusters at redshift $z=0.03$ (left column) and at redshift $z=0.4$ (right column).  
  The red lines 
  indicate $P_{\gamma\to\gamma}=1.8\%$.
  } 
  \label{OE} 
\end{figure}

In Fig.\ (\ref{OE}), 
we further compare the theoretical upper bound on the photon 
survival probability in the $z \geq z_0$ region, 
which is $P_{\gamma\to\gamma} < 1.8\%$, to the actual calculations given in Ref.~\cite{Galanti:2022yxn}, for photons produced in galaxy clusters and propagating to Earth, 
where  different medium effects and 
varying magnetic fields are taken into consideration. 
We find that in the energy range where the photon 
polarization $\Pi_L \approx 1$, 
which is $\omega \gtrsim 2 \times 10^4$ GeV ($\omega \gtrsim 10^3$ GeV) 
on the left (right) panel figure of Fig.\ (\ref{OE}), 
the photon survival probability is indeed below $1.8\%$.

\section{Propagation modes for photons with energy $\omega = 10^{-3}-10^2$ GeV}
\label{sec:lowE}

In this section, we discuss the propagation modes 
for photons with energy $\omega = 10^{-3}-10^2$ GeV. 
In this energy range, 
as compared to 
the ALP-photon mixing term, 
the quantity $\Gamma$ can also be considered negligible,  
in addition to the $\Delta_{x}$ and $\Delta_{y}$ terms, 
which are considered to be negligible in section \ref{TWO}.  
Thus our analytical analysis carried out in 
section \ref{TWO} is still valid and can be 
further simplified by taking the $\Gamma \to 0$ limit. 
In section \ref{TWO} we have neglected the ALP mass term. 
In this section we also study the effects of the 
ALP mass term.

We first discuss the case where the ALP mass term is neglected. 
In this case, we take the $\Gamma \to 0$ limit in 
Eqs.~(\ref{AX1}) and (\ref{AY1}), which lead to 
\begin{align}
\label{eq:massless:rhox}
&\rho_x(z)=\frac{1}{2}, \\
&\rho_y(z)= \frac{1}{2} \cos^2 \left( \frac{g_{a\gamma}Bz}{2} \right). 
\label{eq:massless:rhoy}
\end{align}
Thus, both $\rho_x(z)$ and $\rho_y(z)$ are independent on energy, 
which also result in energy-independence of the photon survival 
probability $P_{\gamma \to \gamma }$ and 
the photon degree of polarization $\Pi_{L}$.

\begin{figure}[t]
  \centering
  \includegraphics[width=0.5\textwidth]{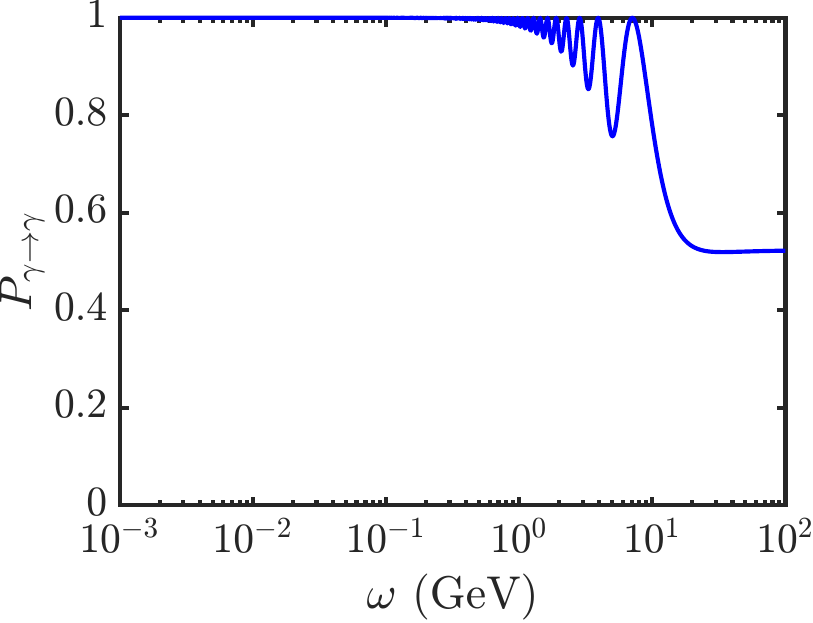} 
  \caption{Photon survival probability as a function of the energy. 
  Here we take the magnetic field as $B=5\ \mu\text{G}$, 
  the coupling constant as $g_{a\gamma}=3\times 10^{-12}$ GeV$^{-1}$, 
  the propagation distance as $z=1\ \text{Mpc}$, 
  and the ALP mass as $m_a=10^{-18}\ \text{GeV}$.} 
  \label{BD1} 
\end{figure}

We next discuss the case where the ALP mass term is important. 
We consider the case where $m_a=10^{-18}\ \text{GeV}$.
In this case, we have $\rho_x(z)=1/2$. 
The analytical form of $\rho_y(z)$ can be obtained by replacing 
$\Gamma$ in Eq.\ (\ref{AY1}) with $2i\Delta_{aa}=-im_a^2/\omega$.
Thus, we have 
\begin{equation}
\label{AYD4}
\rho_y(z)=\frac{1}{2}\left|\cos \left(\frac{\Delta}{2}z\right)+i\frac{\Delta_{aa}}{\Delta}\sin \left(\frac{\Delta}{2}z\right)\right|^2,
\end{equation}
where $\Delta=\sqrt{g_{a\gamma}^2B^2+m_a^4/(4\omega^2)}$. 
We further compute the photon survival probability $P_{\gamma\to\gamma}$ and 
the polarization degree $\Pi_{L}$ 
by $P_{\gamma\to\gamma}=1/2+\rho_y$ and 
\begin{equation}
\label{eq:relation}
\Pi_{L}=\frac{1-2\rho_y}{1+2\rho_y}
= (P_{\gamma\to\gamma})^{-1}-1. 
\end{equation} 
Thus, the polarization degree is solely determined by 
the photon survival probability $P_{\gamma\to\gamma}$. 
We note that the relationship between the polarization degree of photons 
and their survival probability has been discussed in Refs.~\cite{Galanti:2022ijh,Galanti:2022yxn,Galanti:2023uam,Galanti:2022iwb,Galanti:2022tow}. However, these references focused on the connection between the survival probability 
and the initial polarization degree. 
In contrast, Eq.\ (\ref{eq:relation}) 
addresses the relationship between the survival probability 
and the polarization degree at the time of observation.

Fig.\ (\ref{BD1}) shows 
the relation between the photon survival probability $P_{\gamma\to\gamma}$ 
and the photon energy $\omega$. 
Interestingly, we find that there exist two 
energy regions where the photon survival probability 
$P_{\gamma\to\gamma}$ is (nearly) independent of the photon energy: 
(1) $\omega \lesssim 1$ GeV, 
(2) $\omega \gtrsim 10$ GeV.
This can be understood as follows. 
If the photon energy $\omega$ is high such that 
the term $\Delta_{aa}=-m_a^2/(2\omega)$ 
can be neglected, this is just the case 
where the ALP mass can be neglected. 
Thus one should have energy-independent 
$\rho_x$ and $\rho_y$, as given in 
Eqs.~\eqref{eq:massless:rhox} 
and \eqref{eq:massless:rhoy}, 
resulting in energy-independence in $P_{\gamma\to\gamma}$ 
for $\omega \gtrsim 10$ GeV in Fig.\ (\ref{BD1}). 
If the photon energy is low such that 
$\Delta_{aa}/\Delta \to 1$, 
we have $\rho_y\to1/2$. 
This then leads to $P_{\gamma\to\gamma} \approx 1$ 
for $\omega \lesssim 1$ GeV in Fig.\ (\ref{BD1}).

The measurement of photon polarization degree can be carried out in the energy range of $10^{-3}-10^{1}\ \text{GeV}$ through advanced gamma-ray detectors, such as COSI\ \cite{COSI}, e-ASTROGAM\ \cite{Kawata:2017zyo,e-ASTROGAM:2018jlu}, and AMEGO\ \cite{Kierans:2020otl}, in the upcoming future.
We note that the polarization degree, $\Pi_{L}$, can be indirectly determined using the relation $\Pi_{L}=(1/P_{\gamma\to\gamma}-1)$.
Additionally, due to the photon-ALP interaction, 
the photon energy spectrum can exhibit a distinct oscillatory pattern 
sandwiched between two almost energy-independent regions, 
as illustrated in Fig.\ (\ref{BD1}), 
which may serve as a novel signature for the ALP detection.

\section{Conclusions}
\label{sec:conclusions}

In this paper we have identified two distinct photon propagation modes in the presence of ALPs. We classify the two different modes by the sign of $D= (g_{a \gamma} B)^2 - \Gamma^2/4$. For the $D>0$ propagation mode, the intensity of photon oscillates as propagating, producing multiple peaks along its propagation path. For the $D<0$ case, on the contrary, the intensity of photon does not exhibit any oscillatory behavior.

We use analytic methods to study the two photon propagation modes, because they are not readily discernible in numerical simulations, which have been extensively used in the literature to model the photon-ALP propagation.  
In our analytic methods, we assume a uniform magnetic field and negligible medium effects so that the propagation equation of the photon-ALP system can be solved 
in a simple analytic form.

We investigate photon propagation in two energy regions where our analytic methods are appropriate: (1) photon with energy $\omega > 100$ GeV; (2) photon in the energy range of $10^{-3}-10^2$ GeV. We identify the two photon propagation modes by comparing the magnetic field and the photon attenuation rate $\Gamma$ in different astrophysical environments. For the two propagation modes, We compute the photon survival probability $P_{\gamma \to \gamma}$ and the degree of photon polarization $\Pi_L$.

In the $D>0$ propagation mode, 
the fully-polarization can occur 
either in the $\rho_x\ll \rho_y$ case or in the $\rho_y\ll \rho_x$ case. 
Because of the oscillatory behavior in the intensity, 
the fully-polarization exhibits as 
various peaks in the propagation distance. 
In the $D<0$ propagation mode, 
there is no oscillation in the photon intensity. 
The detection condition that yields optimal results should include both a nearly full-polarization signal and a considerable photon survival probability. 
The distances at which this condition is met in the $D<0$ case 
are $z \simeq z_0$ and $z \simeq 2z_0$, 
where $z_0$ is defined in Eq.~\eqref{z0s}.
We further find an upper bound on the photon survival probability, 
$\simeq 1.8$ \%, for the full-polarization region, $\Pi_L \approx 1$.

In the energy interval of $10^{-3}< \omega < 10^{2}$ GeV, 
both medium and attenuation effects can become 
small compared to the ALP-photon mixing term, 
leading to even simpler analytic results. 
In this energy range, the propagation mode is predominately 
the $D>0$ mode in most of the parameter space of interest. 
We further find some distinguishing signatures associated 
with the ALP mass $m_a$: 
If $m_a\simeq0$, 
both $P_{\gamma\to\gamma}$ and $\Pi_{L}$ 
are energy-independent; 
If $m_a$ cannot be neglected, there exist two 
energy-independent regions separated by an oscillating pattern 
in the energy spectrum of $P_{\gamma \to \gamma}$ and $\Pi_{L}$, 
which may serve as a novel signature to 
detect ALPs in the future experiments.

\section{Acknowledgements}

The work is supported in part by the 
National Natural Science Foundation of China under Grant Nos.\ 
11675002, 11725520, 12147103, 12235001, and 12275128. 
ZL thank Yonglin Li and Zi-Wei Tang for discussions.

\appendix

\section{Comparison between analytic and numerical calculations}
\label{app:Appendix}

In this section we compare analytic calculations 
based on a uniform magnetic field assumption, 
with the numerical calculations using 
a more realistic consideration of the 
magnetic field, both in the MW galaxy 
and in galaxy clusters.

\subsection{MW galaxy}

We first discuss the photon propagation in the MW galaxy.
We adopt the MW magnetic field model given in 
Refs.~\cite{Jansson:2012pc,Jansson:2012rt}, 
in which the magnetic field consists of two components: 
the regular component 
and the random component. 
Following Ref.~\cite{Bi:2020ths}, 
we neglect the random fields in our analysis. 
For completeness, we present the regular component of 
the MW magnetic field from Refs.~\cite{Jansson:2012pc,Jansson:2012rt} 
in Table \ref{MKM}.

\begin{table}[htbp]
\centering
\begin{tabular}{|c|cccccccc|}

\hline

      $i$ & 0 & 1 & 2 & 3 & 4 & 5 & 6 & 7\\
\hline
   $r_{xi}$ (kpc) & 5.1  &  6.3 & 7.1 & 8.3 & 9.8 & 11.4 & 12.7 & 15.5    \\
\hline
   $b_i$ ($\mu$G)  &  &   0.1 & 3.0 & -0.9 & -0.8 & -2.0 & -4.2 & 0 \\

\hline
\end{tabular}
\caption{The regular component of the 
magnetic field model in the MW disk \cite{Jansson:2012pc}.
The disk is partitioned into 7 regions with 
8 boundaries that are given by 
$r_i=r_{xi} \exp(\phi \tan(90^\circ-\beta))$, 
where $\beta=11.5^{\circ}$; 
the $i$-th region is bounded by 
the two curves $r_{i-1}$ and $r_{i}$. 
The coordinates $(r,\phi)$ are defined such that 
the Galactic center is (0,0) and 
the Sun is (-8.5 kpc, 0). 
The direction of the regular component is 
$\hat{b}=\sin(\beta)\hat{r}+\cos(\beta)\hat{\phi}$; 
the magnitude of the regular component 
in the $i$-th region
is given by 
$B=b_i {r_0}/{r}$ where $r_0=5$ kpc.}
\label{MKM}
\end{table}

\begin{figure}[htbp]
  \centering
  \includegraphics[width=0.45\textwidth]{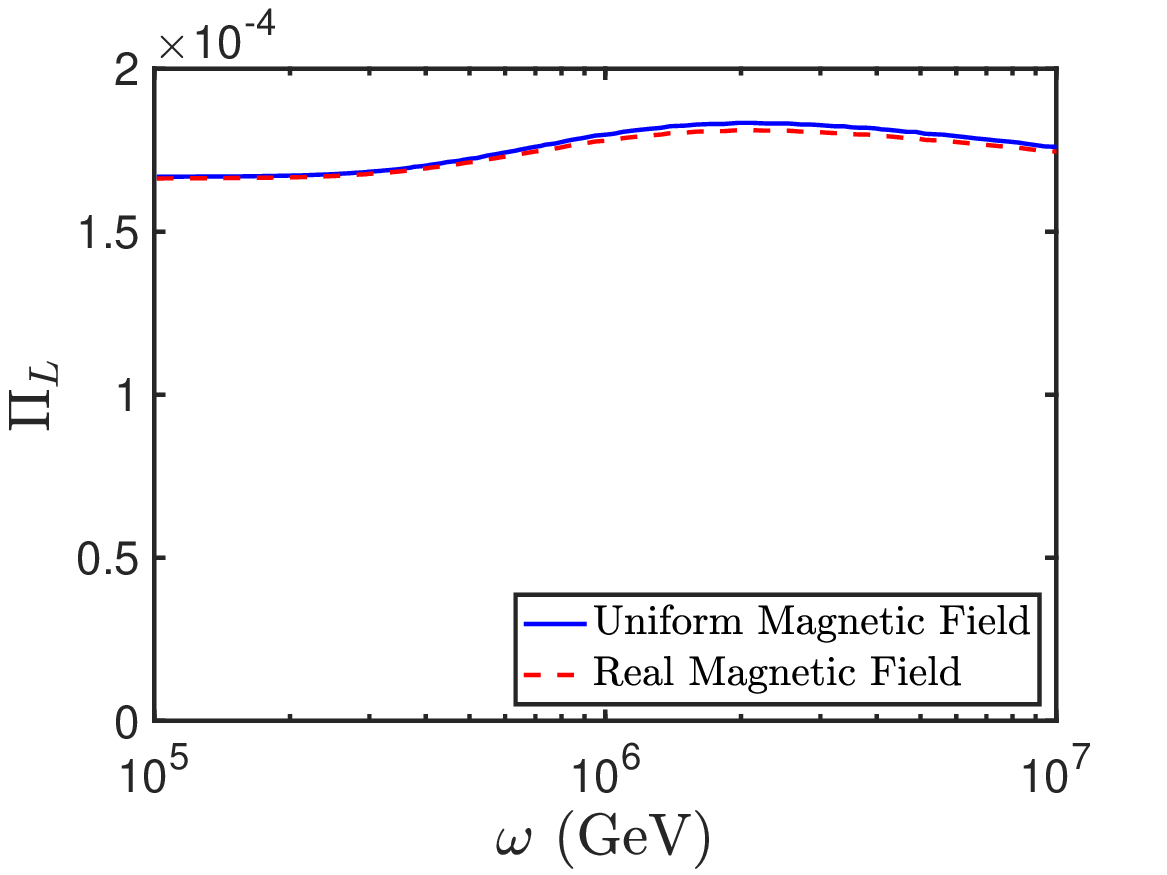}
  \includegraphics[width=0.45\textwidth]{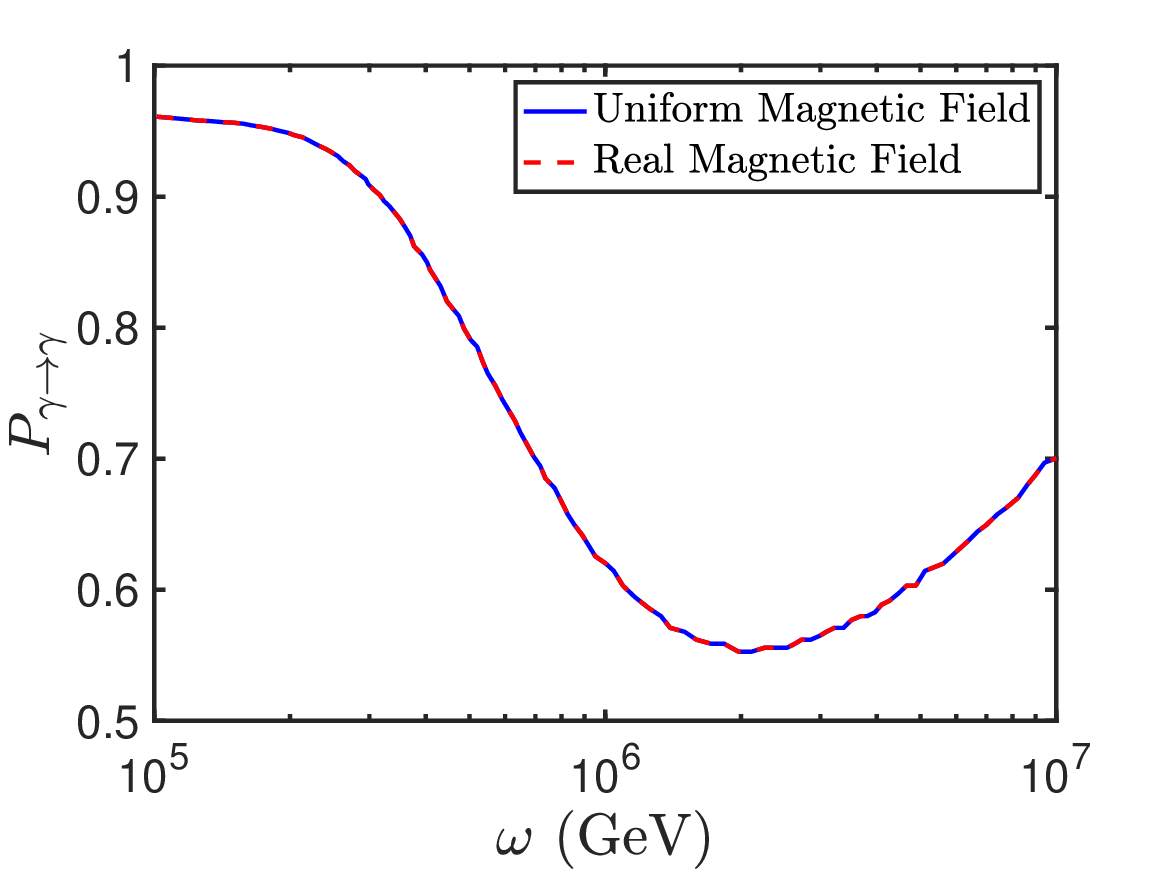}
  \caption{The polarization degree (left) and survival probability 
  (right) as a function of the photon energy in the MW galaxy. 
  The blue lines are obtained in 
  analytic calculations where the medium effects are neglected and 
  the magnetic field $B_0=0.98\ \mu$G is used. 
  The red lines are obtained in 
  numerical calculations
  where the medium effects ($\Delta_x$ and $\Delta_y$) and the MW magnetic field model 
  are used (see main text).}
  \label{MK} 
\end{figure}

We consider a source in the direction of 
$(\ell,b) = (180^\circ,0^\circ)$ 
(the opposite direction of GC) 
and with a distance of 4 kpc from Earth. 
In our analytic calculation
we use $B_0=0.98\ \mu$G,
which is the average value of the magnitude of the 
magnetic field along the propagation path.
Fig.~(\ref{MK}) shows the polarization degree and survival probability, 
computed both in the analytic calculations 
and in the numerical calculations 
where the medium effects and the MW magnetic 
field model are used. 
The agreements between the analytic and numerical 
calculations indicate that the analytic calculation 
with a uniform magnetic field is a good approximation.

\begin{figure}[htbp]
  \centering
  \includegraphics[width=0.45\textwidth]{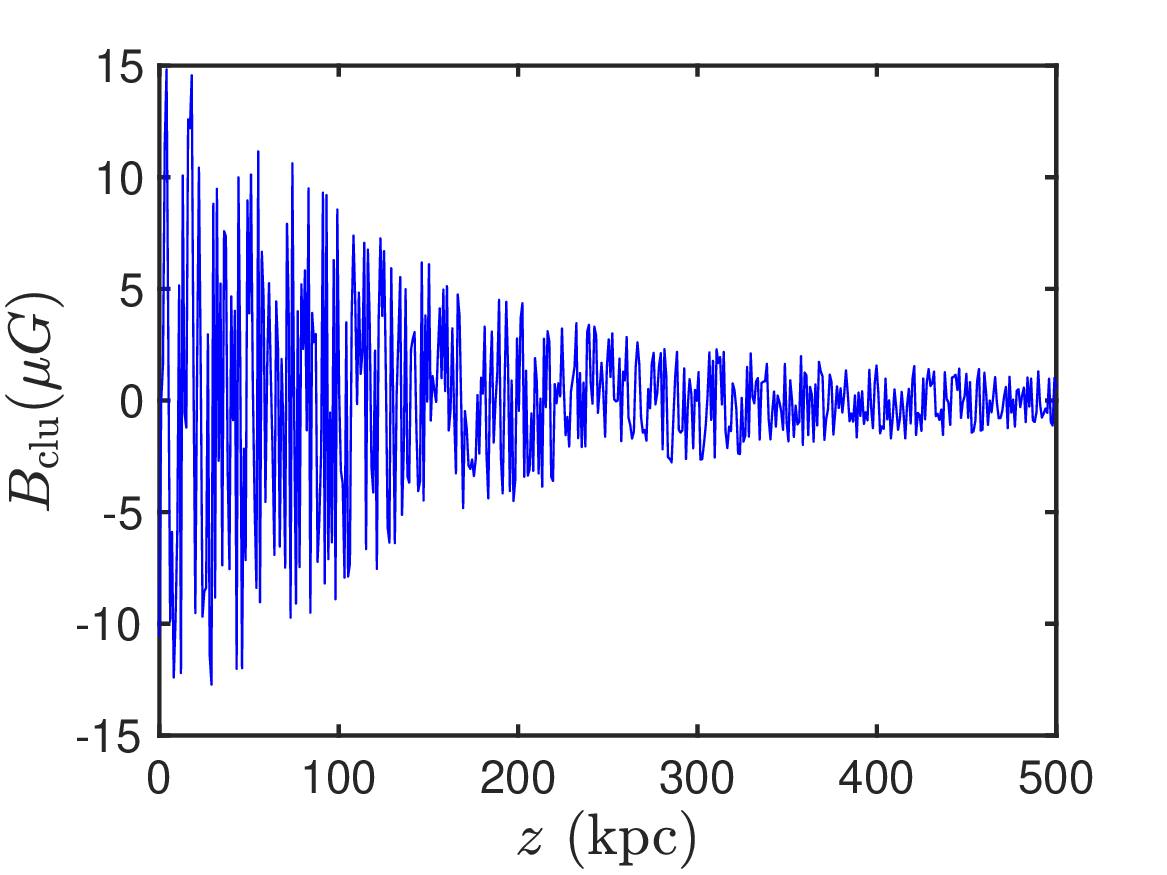}
  \includegraphics[width=0.45\textwidth]{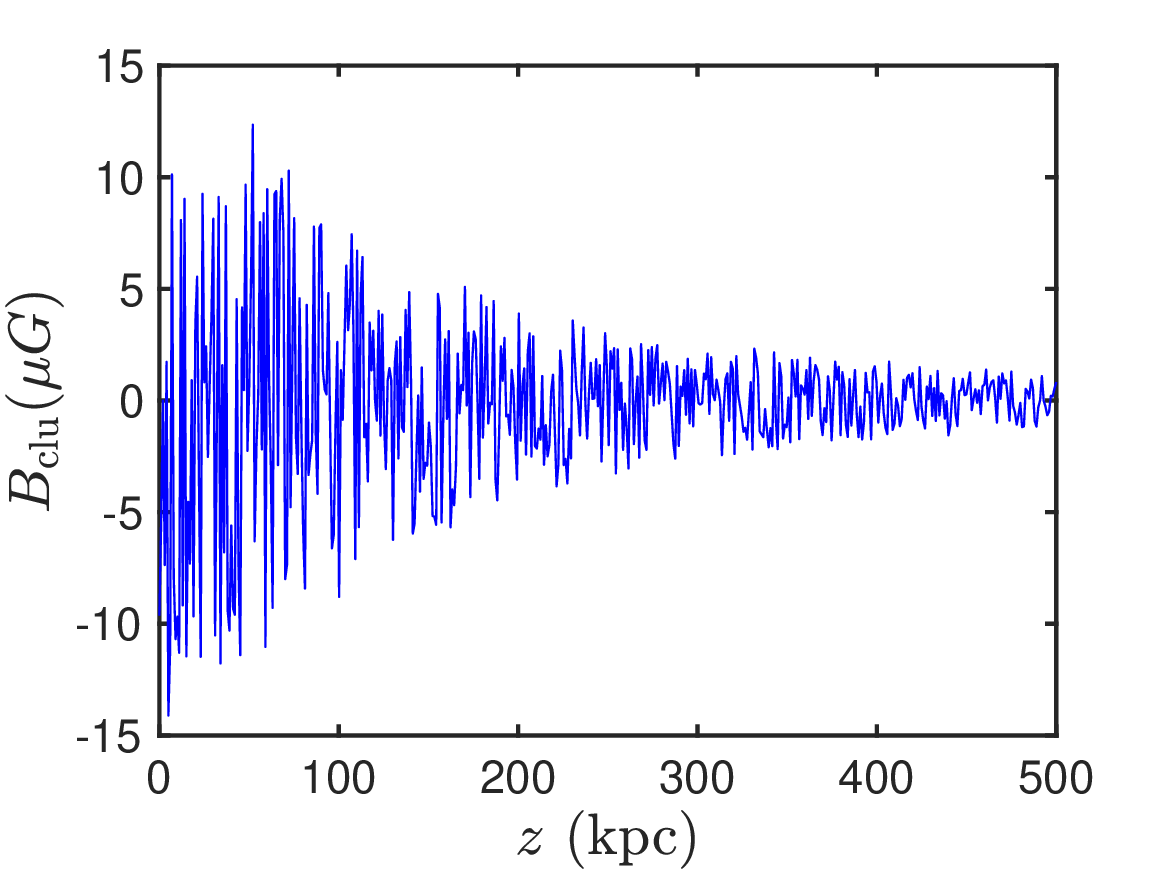}
  \caption{Simulated magnetic fields in the $x$- (left) and $y$- (right) directions in the galaxy cluster as a function of propagation distance $z$.} 
  \label{Bclu} 
\end{figure}

\begin{figure}[htbp]
  \centering
  \includegraphics[width=0.45\textwidth]{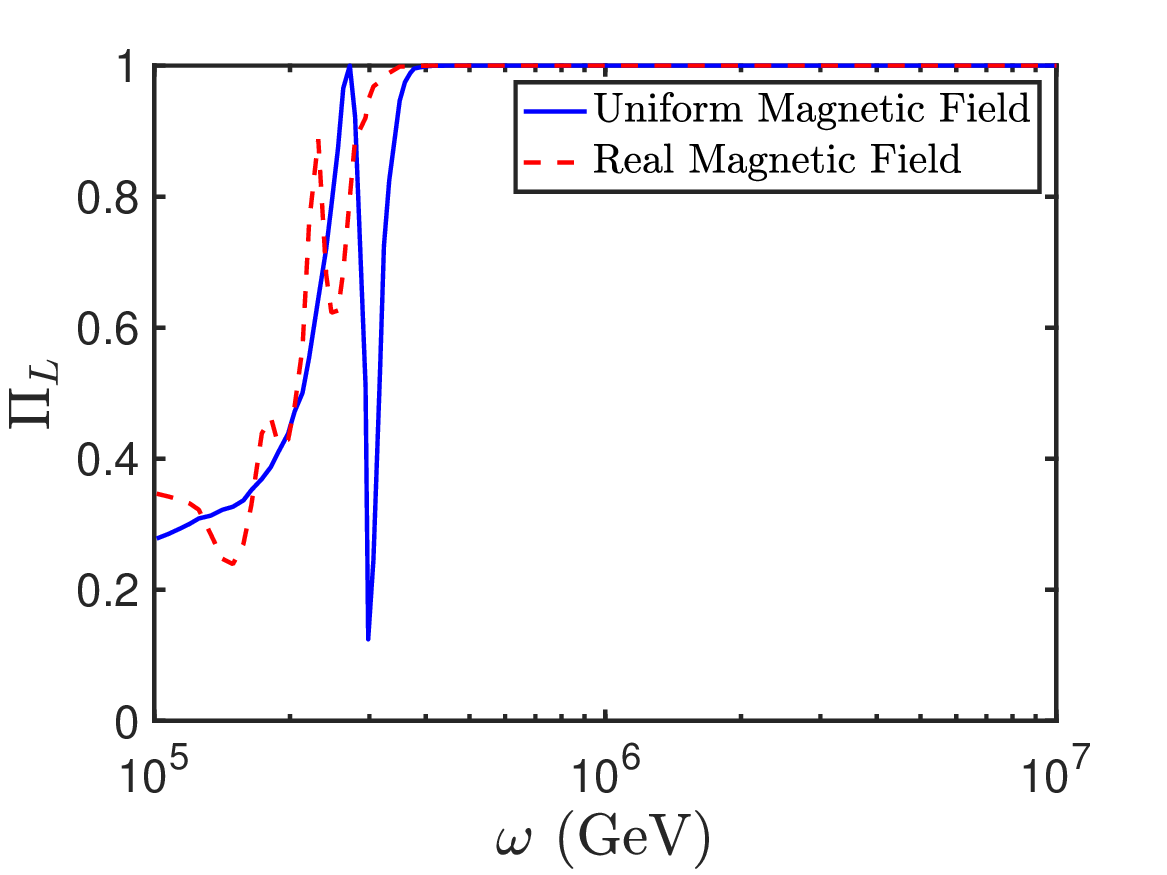}
  \includegraphics[width=0.45\textwidth]{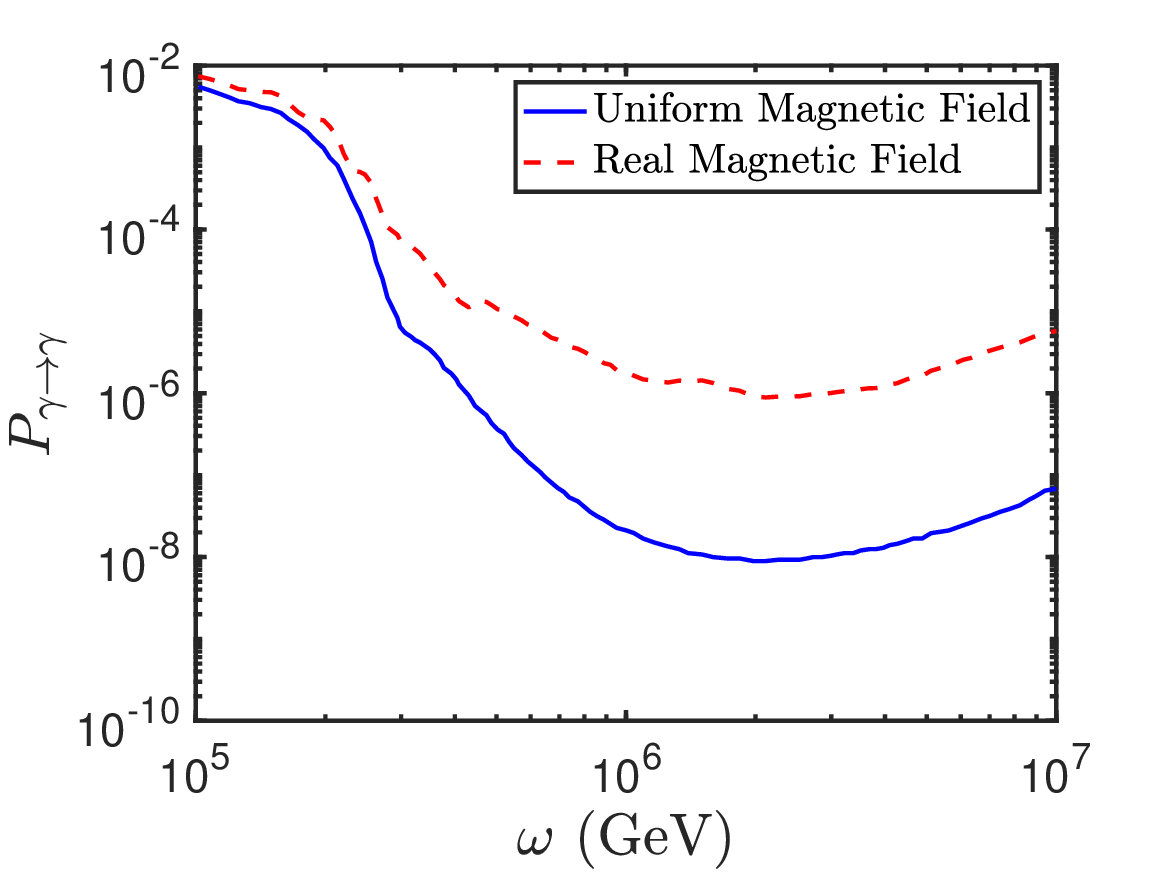}
  \caption{The polarization degree (left) and survival probability 
  (right) as a function of the photon energy in the a galaxy cluster. 
  The blue lines are obtained in 
  analytic calculations where the medium effects are neglected and 
  a uniform magnetic field of $B_0=0.19\ \mu$G is used. 
  The red lines are obtained in 
  numerical calculations
  where the medium effects are considered 
  and the magnetic fields shown in Fig.~\ref{Bclu} 
  are used.}
  \label{GC} 
\end{figure}

\subsection{Galaxy clusters}

We next discuss galaxy clusters. 
To model the complex magnetic field in galaxy clusters, 
we use the parametrization given in Refs.~\cite{Galanti:2022yxn,Bonafede:2010xg}
\begin{equation}
\label{BcluS}
B_{\rm clu}(z)=B_{\rm clu 0}
\left( 1+\frac{z^2}{r_{\rm core}^2} 
\right)^{-\frac{3}{2}\eta \beta}. 
\end{equation}
For galaxy clusters, we adopt the following parameters:
$B_{\rm clu0}=15$ $\mu$G, $r_{\rm core}=100$ kpc, $\beta=2/3$, and $\eta=0.75$ \cite{Galanti:2022yxn}. 
Following Ref.~\cite{Bonafede:2010xg}, 
we assume that the direction of the 
magnetic field is random; 
hence, we simulate the magnetic field direction by 
using Monte Carlo method with a step of 1 kpc along 
the propagation path. 
Fig.~(\ref{Bclu}) shows the simulated magnetic fields in 
the $x$- (left) and $y$- (right) directions in the galaxy cluster as a function of propagation distance $z$.

We obtain the mean value of the magnetic field in the galaxy cluster 
by taking the average of the magnetic field along the propagation 
path. 
In our simulation, we obtain $B_0=0.19\ \mu$G, 
which is the average of the magnetic field 
from $z=0$ to $z=500$ kpc. 
Fig.~(\ref{GC}) shows the polarization degree and survival probability, 
computed both in the analytic calculations 
and in the numerical calculations 
where the medium effects are considered 
and the magnetic 
field model given in Eq.~\eqref{BcluS} is used. 
We note that although the analytic calculation 
deviates from the numerical calculation, 
it describes the behavior qualitatively 
and provides valuable insights for ALP effects.

\normalem

\bibliography{ref.bib}{}
\bibliographystyle{utphys28mod}

\end{document}